\documentclass[twocolumn,showpacs,floatfix,aps]{revtex4}
\usepackage[dvips]{graphics}

\def \beq {\begin{equation}}
\def \eeq {\end{equation}}

\usepackage{amssymb}

\begin{document}

\title{Gravitational wave recoil in Robinson-Trautman spacetimes }

\author{Rodrigo P. Macedo}
\email{romacedo@fma.if.usp.br}
\affiliation{
Instituto de F\'\i sica,   Universidade de S\~ao Paulo, \\
C.P. 66318, 05315-970   S\~ao Paulo, SP, Brazil.}
\author{Alberto Saa}
\email{asaa@ime.unicamp.br}
\affiliation{
Departamento de Matem\'atica Aplicada,   Universidade Estadual de Campinas, \\
C.P. 6065, 13083-859 Campinas, SP, Brazil.}

\begin{abstract}
We consider the gravitational recoil due to non-reflection-symmetric  gravitational wave
emission in the context of axisymmetric Robinson-Trautman spacetimes. We show that regular
initial data evolve generically into a final configuration corresponding to a
 Schwarzschild black-hole moving with constant speed.
 For the case of (reflection-)symmetric initial  configurations, the mass of the remnant black-hole and the
 total energy radiated away are completely determined by the initial data, allowing us to
 obtain analytical expressions for some recent numerical results
 that have been appeared in the  literature. Moreover,
by  using the Galerkin spectral method to analyze  the   non-linear regime
 of the Robinson-Trautman equations, we show that
the recoil velocity can be estimated with good accuracy  from
 some  asymmetry measures (namely the first odd moments) of the initial data.
 The extension for the non-axisymmetric case and
the  implications of our results for realistic situations  involving head-on collision
of two black holes are also discussed.
\end{abstract}

\maketitle
\section{Introduction}
The possibility that a body recoils while emitting gravitational radiation has been known
for decades\cite{recoil}. This problem has been considered in the literature by means of
 many approximated and
semi-analytical methods as,
for instance, the particle approximation\cite{particle}, post-Newtonian methods\cite{postNewt},
and the Close-Limit Approximation\cite{closed}, leading to typical recoil velocities of
few hundreds of km/s  for some realistic cases. Such conclusions, however,  have
changed drastically
due to some recent advances in numerical relativity\cite{numrel}. In particular,
recent numerical simulations\cite{merger} of the merging process of binary black-holes    indicate that asymmetrical gravitational wave
emission can indeed induce the merger remnant to recoil with velocities up to several
thousands of km/s.
The physical nature and possible implications of such considerably higher
gravitational recoil are now under intense investigation (see, for instance, \cite{consequences}).
The calculation of the recoil velocity as a function of the black-holes initial  conditions is a
particularly important hard task. Since the full non-linear regime  of Einstein equations is
extremely intricate and costly to analyze,   some approximated or ``empirical'' formulas relating
 the  recoil velocity and the initial data  have been proposed\cite{Kick}.

 The Robinson-Trautman (RT) spacetime \cite{RT} is perhaps the simplest
solutions of General Relativity which can be interpreted as an  isolated gravitational
radiating system  and, hence, it  is certainly pertinent to the study of the gravitational
recoil effect. However, despite the many strong mathematical results on the RT solutions available
in the literature, only a   few exact examples of RT spacetime are indeed known
in explicit form (see, for references, \cite{book}).
It is known, nevertheless, that a regular initial data, corresponding typically
to a compact body surrounded by gravitational waves, will evolve smoothly according to
the RT equation into a final
state corresponding to a remnant Schwarzschild black-hole\cite{Chrusciel}, which
can be at rest or moving with constant speed.
Our aim here is to go a step further in the  characterization of such final evolution state as
function of the initial conditions. Our results are motivated and checked by some numerical
analysis.
The Robinson-Trautman partial differential equation
has been analyzed numerically in the recent literature\cite{RTnum}, being
   particularly  suitable to be numerically solved
by means of spectral methods\cite{spectral,rio,rio1}. We will follow Oliveira and Dami\~ao Soares\cite{rio,rio1}
and adopt the Galerkin method\cite{galerkin} for our analysis. However, as we will show, we will implement
it in a different way that will allow us to get simpler equations and a better accuracy.

 The present paper has four Sections and one Appendix.
In the next Section, the main aspects of axisymmetric RT spacetimes are  presented briefly. We show,
in particular,
 how to read from the final state of the RT evolution the mass and speed of the remnant
black-hole. It is also shown that, as expected,
  for (reflection-)symmetric initial data there is no radiation recoil.
In such a case,
the mass of the remnant black-hole and the
 total energy radiated away are completely determined by the initial data, allowing  us to
 establish analytical expressions for the results about the total radiated energy
 obtained numerically in \cite{rio} and \cite{rio1}.
 Section III is devoted to the study of generic axisymmetric initial data. We show that a typical RT
 evolution can  lead to a gravitational recoil. A Galerkin projection method is used to calculate the
 final black-hole speed. We show also how the final recoil velocity can be estimated with good
 accuracy from some
asymmetry measures of the initial configuration, namely the first odd moments of the initial data.
In the
last Section, we discuss the
physical interpretation of the typical initial data considered in this work, emphasizing
their relation with the problem of frontal collision of two black-holes. The extension of our results
to the non-axisymmetric case is also commented in the last section.
 The Appendix presents a direct proof of a mathematical result used in Section II,
 namely that, for regular
initial data, the final state of the
RT evolution does correspond generically to a Schwarzschild black hole moving with constant speed.

\section{Axisymmetric RT spacetime}
The standard form of the
Robinson-Trautman (RT) metric in the usual spherical radiation coordinates $(u,r,\theta,\phi)$ reads\cite{book}
\beq
\label{RT}
ds^2 = -\left(K  - 2\frac{m_0}{r}  - r (\ln Q^2)_u\right)du^2 - 2dudr +\frac{r^2}{Q^2}d\Omega^2,
\eeq
where $Q=Q(u,\theta,\phi)$, $m_0$ is a constant mass parameter, and
$d\Omega^2$ and $K$ stand for, respectively, the metric of the unit sphere  and
the gaussian curvature   of the surface  corresponding to $r=1$ and $u=u_0$ constant, which is given by
\beq
K = Q^2\left(1 + \frac{1}{2}\nabla^2_\Omega \ln Q^2 \right),
\eeq
with $\nabla^2_\Omega$ corresponding to the Laplacian on the unit sphere.
Vacuum Einstein's equations for the metric (\ref{RT}) implies the Robinson-Trautman non-linear
partial differential equation\cite{book}
\beq
6m_0\frac{\partial}{\partial u} \left( \frac{1}{Q^2}\right) = \nabla^2_\Omega K.
\eeq
In this paper,
we will focus on axisymmetric spacetimes and hence we will assume hereafter that $Q=Q(u,\theta)$. By
introducing $x=\cos\theta$   one has
\beq
\label{lambda}
K = Q^2 + Q \frac{\partial }{\partial x} \left[(1-x^2) Q_x \right] - (1-x^2)Q_x^2
\eeq
and
\beq
\label{RT1}
6m_0\frac{\partial}{\partial u} \left( \frac{1}{Q^2}\right) = \left[(1-x^2)K_x \right]_x,
\eeq
where
\begin{eqnarray}
\label{lambdax}
\left[(1-x^2)K_x \right]_x &=& (1-x^2)^2\left( QQ_{xxxx} - Q_{xx}^2 \right)  \\
&-&   8(x-x^3)QQ_{xxx}
 - 4(1-3x^2)QQ_{xx}. \nonumber
\end{eqnarray}
Integrating (\ref{RT1}) and assuming a regular gaussian curvature $K$
one has
\beq
\label{constr}
\frac{d}{d u}\int_{-1}^1 \frac{dx}{Q^2(u,x)} = 0,
\eeq
implying that the quantity $q_0 = \int_{-1}^1 Q^{-2}  dx  $ is  constant
 along the solutions of (\ref{RT1}). Notice that, from (\ref{RT}), the regularity of the surface
 $u$ and $r$ constants precludes us of having $Q=0$.
 The regularity of the gaussian curvature $K$, on the other hand,
  requires $0<Q<\infty$.
 We normalize our data in order to have $q_0=2$,
 implying that the area of the surface corresponding to $r$ and $u$ constants is  always
 $4\pi r^2$   along the
 $u$-evolution governed by  (\ref{RT1}).

Several classical results assure that, given a geometrically regular   initial
data $Q(0,x)$, the solution of (\ref{RT1}) approaches asymptotically a stationary ($Q_u=0$) regime.
The stationary  solutions of (\ref{RT1}) are such that
\beq
(1-x^2)K_x = A = {\rm\ constant},
\eeq
leading to
\beq
K = A\,{\rm  arctanh}\, x + B,
\eeq
where $B$ is another constant. Regularity of $K$ on the interval $[-1,1]$
requires necessarily $A=0$. On the other hand,
  Eq. (\ref{lambda}) implies that the regular $Q$ solutions for which $K$ is constant
are such that $Q_{xx}=0$ (see the Appendix for a direct proof). Therefore,
the stationary solutions of (\ref{RT1}) are always   of the form
$Q = a + bx$  with $a$ and $b$ constants. Nevertheless, our choice of $q_0=2$ yields
$a^2-b^2=1$. We choose in this work
a parametrization such that $a=\cosh \alpha$ and $b=\sinh \alpha$.

Given a normalized regular
 initial data $Q(0,x)$, the asymptotic   solution of (\ref{RT1}) will be always of the form
$Q(\infty,x)= \cosh\alpha + x\sinh\alpha$.
The final configuration is, hence, completely characterized by the sole parameter $\alpha$.
In order to unveil its physical role,
let us consider the Bondi's mass function\cite{mass}
\beq
\label{mass}
M(u) = \frac{m_0}{2}\int_{-1}^{1} \frac{dx}{Q^3(u,x)}
\eeq
which
has several desirable properties to define an ``instantaneous''  mass for the
solutions of (\ref{RT1}), see, for instance, \cite{rio}.
In particular, we have that $M(u) \ge m_0$ for normalized initial data and, for $u\rightarrow\infty$, it reduces to
\begin{eqnarray}
M(\infty) &=& \frac{m_0}{2}\int_{-1}^1 \frac{dx}{(\cosh\alpha + x\sinh\alpha )^3} = m_0\cosh\alpha \nonumber \\
&& \quad\quad\quad\quad\quad\quad\quad\quad\quad\quad\quad = \frac{m_0}{\sqrt{1-v^2}},
\end{eqnarray}
where   $v=\tanh\alpha$ can be interpreted as the final velocity along the $z$
axis of the remnant
black-hole\cite{mass}.

The Bondi's mass (\ref{mass}) corresponds to the temporal component of the Bondi's
four-momentum, which for generic (non-axisymmetric) RT solutions is given by\cite{mass}
\beq
P_a(u) = \frac{m_0}{4\pi} \int_{S^2} \frac{\eta^a}{ Q^3(u,\theta,\phi)} dS,
\eeq
where $S^2$ is the unit sphere spanned by the usual coordinates  $\theta$ and $\phi$
and with area element $dS$, and $a=0,1,2,3$, with $\eta^0=1$ and $\eta^i$ being the
 radial
 three-vector directed to the point $(\theta,\phi)$ on the unit  sphere. For axisymmetric configurations,
 the non-vanishing components of the Bondi's four-momentum are $P_0(u) = M(u)$ and
 \beq
 P_3(u) = \frac{m_0}{2}\int_{-1}^{1} \frac{ x}{Q^3(u,x)} dx,
 \eeq
 which corresponds to the momentum carried by the solution along the $z$ axis.
For normalized initial data one has for $u\rightarrow\infty$
\begin{eqnarray}
P_a(\infty)  = \frac{m_0} {\sqrt{1-v^2}}(1,0,0,-v),
\end{eqnarray}
reinforcing  the interpretation of $v$ as the final velocity of the remnant
black-hole.

Notice that, for symmetric (even) initial data $Q(0,x)$, Eq. (\ref{lambdax}) implies that the solutions
$Q(u,x)$
of (\ref{RT1}) are necessarily even for   $u\ge 0$, establishing  that there is no
gravitational recoil ($v=0$) in this case. Such a behavior is, of course, in full agreement with the expectation that gravitation
recoil should be due to non-reflection-symmetric
 gravitational wave emission. Therefore, for even
situations, the
constraint (\ref{constr}) determines completely the final   evolution state.

\subsection{Radiated energy:  reflection-symmetric case}

\begin{figure}[h!]
\resizebox{1\linewidth}{!}{\rotatebox{0}{\includegraphics*{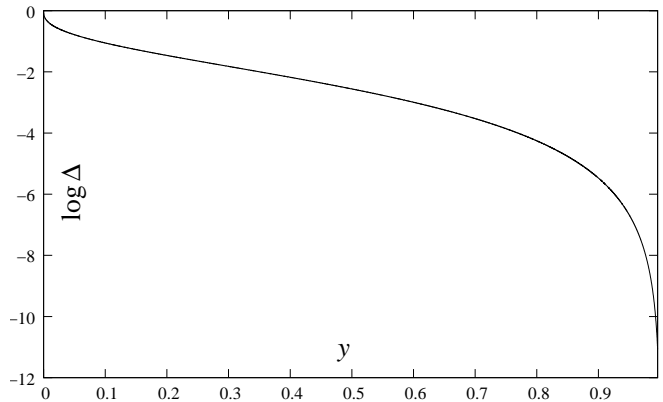}}}
\caption{ The
fraction $\Delta$ of the initial Bondi's mass lost due to gravitational wave
emission for the initial configuration (\ref{prola}), as a function of $y=1-\epsilon$.
 The curve is in very good agreement
with that one inferred from numerical results in \cite{rio,rio1}. Notice, however, that
the non-extensive distribution function proposed in  \cite{rio,rio1} is merely an
approximation for $y\approx 1$, see (\ref{decay}).}
\label{fig1}
\end{figure}

The fraction of the initial mass $M(0)$ radiated away along the $u$-evolution governed by
(\ref{RT1}) can be calculated exactly for even configurations. Following \cite{rio}, we define
\beq
\Delta = \frac{M(0)-M(\infty)}{M(0)},
\eeq
which clearly corresponds to the fraction of the initial mass lost due to gravitational wave
emission. For even configurations, $v=0$ and we have simply
\beq
\Delta = 1 - 2\left( \int_{-1}^{1} \frac{dx}{Q^3(0,x)}\right)^{-1}.
\eeq
It can be shown that $0\le\Delta <1$.
As an explicit example of this exactly soluble case,
let us consider the first even initial data considered in the papers
\cite{rio,rio1}, namely the
prolate spheroid corresponding to
\beq
\label{prola}
Q^2(0,x) = Q_0^2\left( 1-\epsilon^2x^2\right),
\eeq
with $0\le \epsilon < 1$. The constraint $q_0=2$ implies that
\beq
Q_0^2 = \frac{1}{2\epsilon}\ln \left(\frac{1+\epsilon}{1-\epsilon}\right),
\eeq
leading finally to
\beq
\label{delta}
\Delta = 1 -  \sqrt{\frac{1-\epsilon^2}{8\epsilon^3}\ln^3  \left(\frac{1+\epsilon}{1-\epsilon}\right) } .
\eeq
This is the exact analytical expression for the curves obtained in \cite{rio} and \cite{rio1}
 from numerical simulations.
 For sake of
comparison with the results of \cite{rio,rio1}, Fig. \ref{fig1}
depicts a semi-log plot of $\Delta$ as
a function of $y=1-\epsilon$, following their conventions. A very good
agreement is found.
One can proceed in an analogous way for any other even (reflection-symmetric)
configuration, we will return to this issue in the last Section. The exact expression
for $\Delta$ is certainly valuable to the investigation of   statistical properties of the
non-linear gravitational wave emission as those ones considered in \cite{rio,rio1}. For instance,
it is clear from (\ref{delta}) that
the non-extensive
distribution function proposed in  \cite{rio,rio1} is only an
approximation valid for small $\epsilon$.
In fact,   we have
\beq
\label{decay}
\Delta = \frac{1}{30}(1-y)^4 + \frac{32}{945}(1-y)^6 + O((1-y)^8),
\eeq
for $y\approx 1$ (or $\epsilon\approx 0$). Notice that $\Delta\rightarrow 1$ for
$\epsilon\rightarrow 1$.

\section{General Solutions}

\begin{figure}[ht]
\resizebox{1\linewidth}{!}{\rotatebox{0}{\includegraphics*{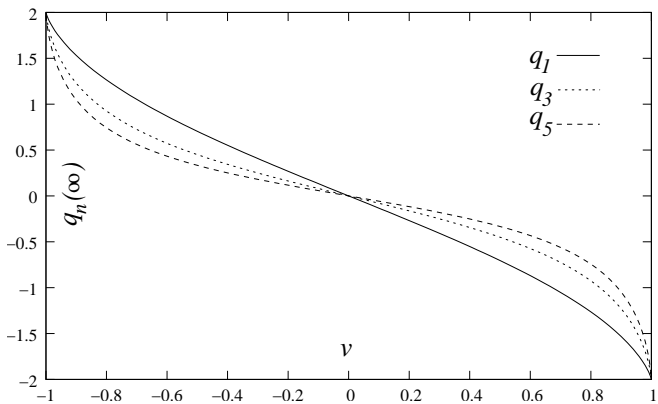}}}
\caption{ Final odd moments $q_n(\infty)$, $n=1, 3, 5$, as  functions of the recoil velocity
$v =\tanh\alpha$, as
given by (\ref{qn}). Notice that, for a given $0 < |v| < 1$, one has $|q_1|>|q_3|>|q_5|>\cdots$}
\label{fig2}
\end{figure}

The evolution of generic initial data $Q(0,x)$ is a greater  challenge. Since the gravitational
recoil is clearly related to the odd part of the function $Q(u,x)$,
 one might consider in first place some asymmetry measures of
the initial data. The simplest ones correspond to their first  odd $n$ moments
\beq
\label{qn0}
q_n(u) = \int_{-1}^{1} \frac{x^n}{Q^2(u,x)} dx,
\eeq
which obey $-q_0 \le q_n \le q_0$.
For the generic final evolution state $Q(\infty,x)= \cosh\alpha + x\sinh\alpha$, we have
\begin{widetext}
\beq
\label{qn}
q_n(\infty) = (1-v^2)\int^{1}_{-1} \frac{x^n}{(1 + vx)^2}dx =
 - \frac{1}{v^n}\left(2-n\frac{1-v^2}{v}\ln\frac{1+v}{1-v}  \right)
  - \frac{1-v^2}{v^{n+1}}\sum_{k=2}^n\sum_{{\rm odd\,}j}^{k-1}\frac{(-1)^k}{k-1}{n \choose k} {k-1 \choose j}v^j,
\eeq
\end{widetext}
valid for odd $n$.
Fig. \ref{fig2}
shows  the first final odd moments $q_n(\infty)$ as   functions of the recoil velocity $v$.
As we will show, the relevance of the first odd moments (\ref{qn0}) rests on the fact that
one can construct, as a linear combination of them, a second approximately conserved quantity
along the solutions of the RT equation (\ref{RT1}) in the framework of the Galerkin
approximation.

\subsection{The Galerkin method}
We introduce now a Galerkin decomposition for $Q(u,x)$
\beq
\label{gal}
Q(u,x) = \sum_{\ell=0}^N b_\ell(u)P_\ell(x),
\eeq
where $P_\ell(x)$ stands for the Legendre polynomials. By using standard projection
techniques\cite{galerkin}, Eq. (\ref{RT1}) can be written as the system of ordinary differential equations
\beq
\label{RTG}
\dot{b}_\ell = -\frac{2\ell+1}{24m_0}\langle Q^3\left[(1-x^2)K_x \right]_x,P_\ell  \rangle,\quad \ell=0,1,\dots,N,
\eeq
where the inner product is given by
$\langle f,g \rangle = \int_{-1}^1 fg \,dx.$ From  (\ref{lambdax}) and (\ref{gal}),
one can see that the functions involved in the inner product in
the right-handed side of (\ref{RTG}) are simple polynomials  in $x$.
The integration can
  be performed  exactly for arbitrary $N$ (with the help of  algebraic
  manipulation software as Maple, for instance),
  yielding $5^{\rm th}$ order polynomials
on the mode functions $b_\ell$. Notice that here, in contrast to the approach adopted in
 \cite{rio,rio1}, no transcendental
function is involved in the Galerkin approximation.
Now, the Cauchy problem for the RT equation
corresponds basically  in choosing the initial value of
the mode functions $b_\ell(u)$ according to
\beq
\label{DATA}
b_\ell(0) = \frac{2\ell + 1}{2 }\langle Q(0,x), P_\ell \rangle,
\eeq
and then to solve the Initial Value Problem (IVP) given by (\ref{RTG}).

Equation (\ref{RTG}) has some useful properties that are independent of $N$.
For instance, their stationary solutions ($\dot{b}_\ell=0$) have necessarily
$b_\ell  = 0$ for $\ell > 1$ and arbitrary (constants)
$b_0$ and $b_1$.
 Indeed, for any regular initial data, the systems evolves into
the final state $Q(\infty,x)= b_0(\infty) P_0(x) + b_1(\infty) P_1(x)$, with
$b_0(\infty)^2 - b_1(\infty)^2 = 1$ for the normalized case,
as expected. The recoil velocity will be given simply by $v=b_1(\infty)/b_0(\infty)$.
Another useful property is that for an even  initial    data, one has
 $b_\ell(u) =0$ for odd $\ell$
and, consequently, $v=0$.
The accuracy of the Galerkin decomposition is determined by the truncation order $N$ in
(\ref{gal}). It can be controlled effectively here by checking
  the conserved quantity $q_0$ along the $u$-evolution.
  Typically, the expansion with $N$ Legendre polynomials in (\ref{gal}) is accurate provided that
$\max |b_N(u)|$ be small enough.

Finally,
we are able now to
consider the evolution of generic initial data. The recoil velocity $v$ can
be calculated by solving the IVP corresponding to
the system of ordinary differential equations (\ref{RTG}) with initial conditions (\ref{DATA}).
The recoil velocity determines  completely the final state for normalized initial data,
allowing the study of any other
relevant quantity as, for instance, the fraction $\Delta$
of the initial mass radiated away as a function
of the non-reflection-symmetric initial  data,
\beq
\label{delta11}
\Delta = 1 - \frac{2}{\sqrt{1-v^2}}\left( \int_{-1}^{1} \frac{dx}{Q^3(0,x)}\right)^{-1} .
\eeq
We have performed an exhaustive numerical analysis of the system
(\ref{RTG}). The considered initial data include  the following simple but
 representative family
\beq
\label{family}
Q(0,x) = Q_0  \left( 1 + \alpha x + \beta x^2 + \gamma x^3\right),
\eeq
where the
  constant  $Q_0$ is always chosen  in order to ensure the normalization $q_0=2$.
  Some particular elements of this family
 \begin{figure*}[ht]
\resizebox{1\linewidth}{!}{ \includegraphics*{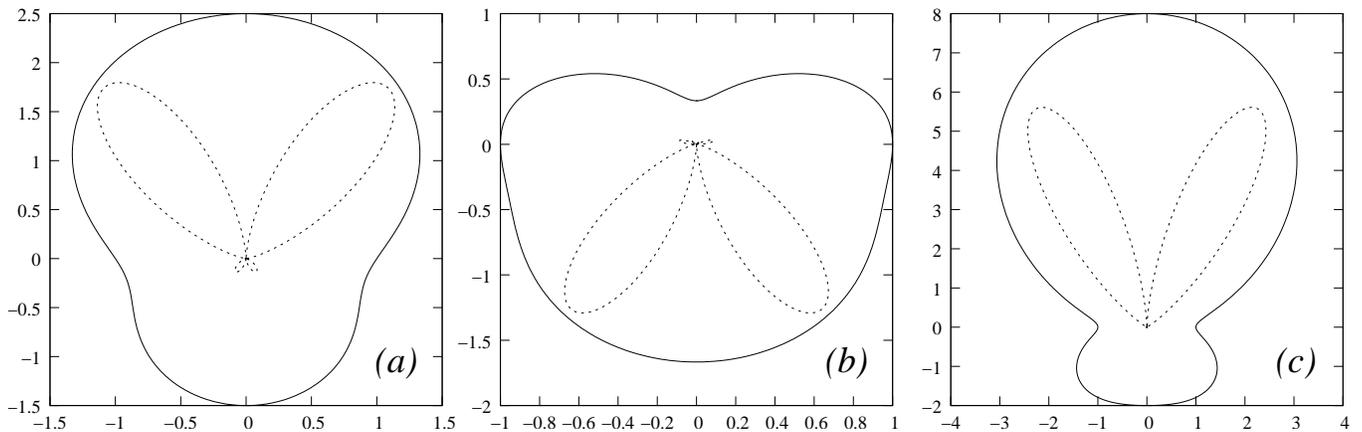}}
\caption{Polar plot of some typical non-reflection-symmetric initial data $Q(0,x)$ of the family (\ref{family}). The initial condition $a$, $b$, and $c$ correspond, respectively, to the parameters $\alpha=1/2$, $\beta=1$, $\gamma=0$;
$\alpha= \beta=0$, $\gamma=-2/3$; and $\alpha=0 $, $\beta=4$, $\gamma=3$. The dashed lines correspond  to
the associated gravitational radiation content (without scale), see Sect. IV.
}
\label{figcond}
\end{figure*}
are presented in Fig. \ref{figcond}.

\begin{figure}[ht]
\resizebox{1\linewidth}{!}{\rotatebox{0}{\includegraphics*{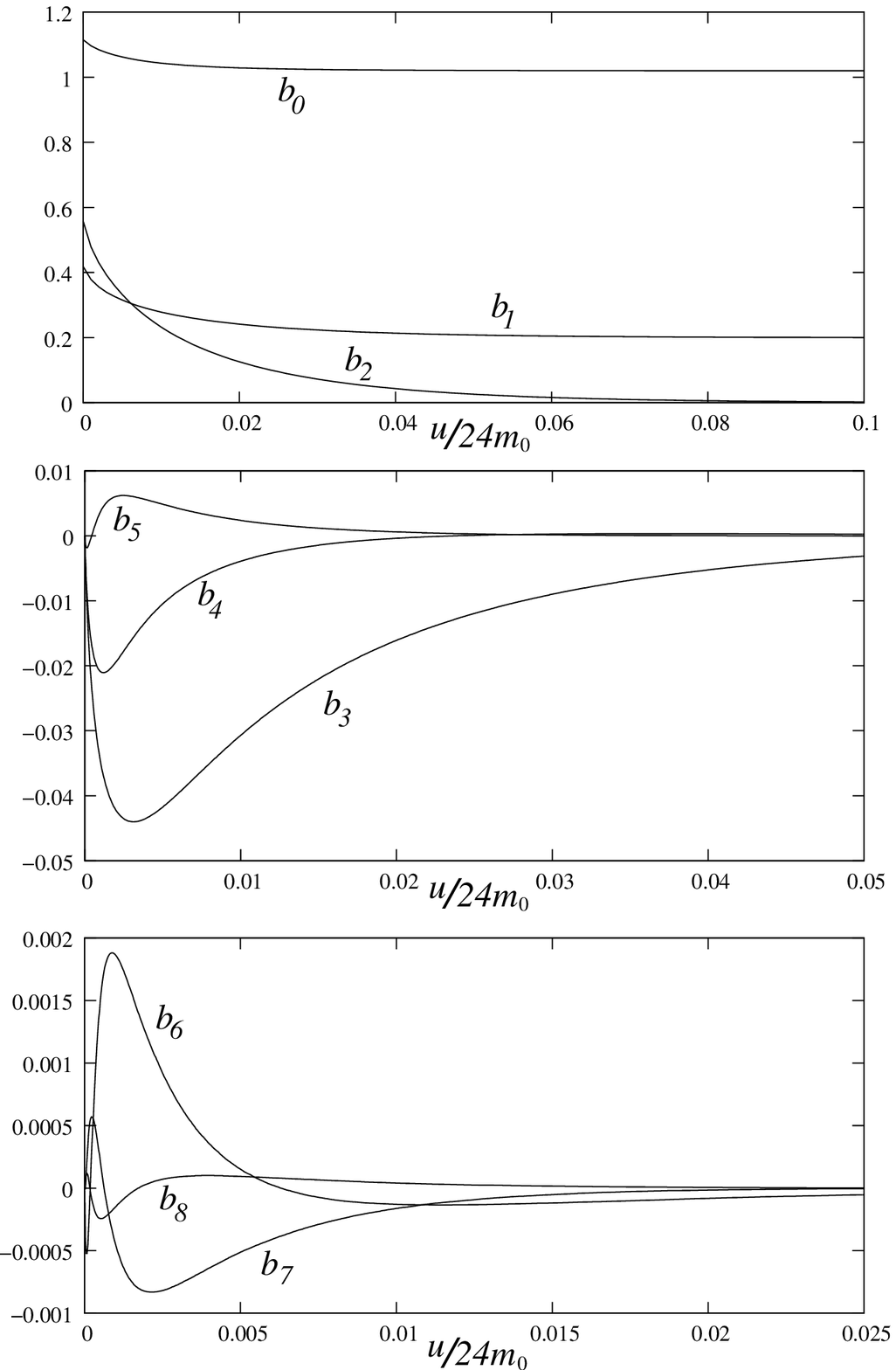}}}
\caption{ Evolution of the modes $b_\ell(u)$ governed by (\ref{RTG}) for the
case $(a)$ of Fig. \ref{figcond}. $N=8$ was used, leading to an accuracy
(controlled by the constant $q_0=2$) of $10^{-4}$. The final evolution state
has $b_0=1.0197$ and $b_1=0.20017$ and, consequently, the recoil velocity is
$v=0.19628$ and the radiated energy fraction $\Delta=0.05420$, calculated
according to (\ref{delta11}).}
\label{figcalc}
\end{figure}

Fig. \ref{figcalc} depicts
a typical evolution for the modes $b_\ell(u)$ governed by (\ref{RTG}) for
a particular case of the family (\ref{family}). The   recoil velocity $v$ can be read from the
final state of the evolution for any initial data.
We notice that, for the family of initial conditions (\ref{family}), we always have $b_\ell(0)=0$
for $\ell>3$  and, in this case,
 $N=8$ is sufficient to assure typically an accuracy (controlled by the constant $q_0(u)=2$) of the Galerkin
 approximation up to 1\%. Initial data with high $q_n(0)$ typically require  higher $N$ in order
 to attain  a given accuracy. The radiation content of the initial data can also give some clues about
 the minimal necessary value of $N$, see   Section IV.

\subsection{Estimation of the recoil  velocity}

\begin{figure}[ht]
\resizebox{1\linewidth}{!}{\rotatebox{0}{\includegraphics*{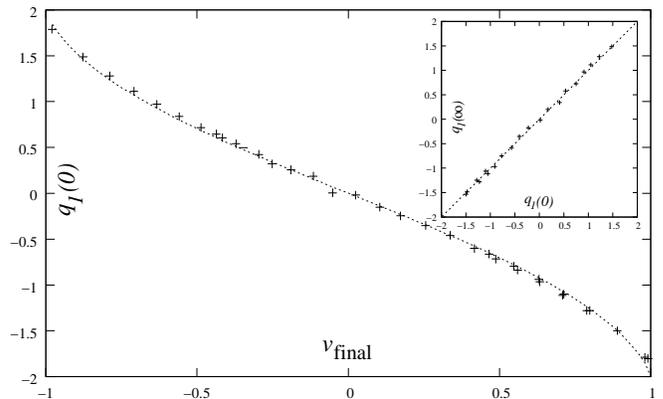}}}
\caption{ Plot of $v\times q_1(0)$ for some typical regular initial conditions.
The dotted line is the curve predicted by (\ref{vfinal}). The detail depicts
the plot  of $q_1(0)\times q_1(\infty)$.
 The assumption of  $q_1(\infty)=q_1(0)$ is, typically,
a good approximation when $q_1(u)$ is   the dominant moment.}
\label{figqn}
\end{figure}

Despite that the IVP associated to the equation (\ref{RTG}) can be solved with
quite modest computational resources, an analytical estimation of the recoil velocity $v$
from the initial data would be certainly valuable. Since the final state of the RT
evolution is completely characterized by the sole parameter $v$ for normalized initial
data, a second conserved quantity
 besides $q_0$ would suffice to determine completely the final state and, consequently, to
 determine the recoil velocity $v$. Unfortunately, the RT
equation (\ref{RT1}) does not seem to have any other conserved quantity   rather than $q_0$.
On the other hand, its Galerkin approximation (\ref{RTG}) does indeed have a second conserved quantity.
Such a new conserved quantity, however, will be only approximately constant along the solutions
of the full RT equation. Nevertheless, the approximation will be as good as the
Galerkin approximation is accurate.
 In order to construct an explicit expression for the
new constant, we remind that (\ref{RT1}) implies that
the moments (\ref{qn0}) obey the equation
\beq
\label{dqn0}
6m_0\dot{q}_n (u) =    \langle x^{n},\left[(1-x^2)K_x \right]_x \rangle.
\eeq
From (\ref{lambdax}) and (\ref{gal}), we see that
\beq
\label{sum}
\left[(1-x^2)K_x \right]_x = \sum_{\ell=0}^{2N} a_\ell(u)x^\ell,
\eeq
where $a_\ell(u)$ are quadratic functions of the modes $b_\ell(u)$. For odd $n$,
the inner product in (\ref{dqn0}) will
select only the odd-$\ell$ terms in $x$ in the summation (\ref{sum}), leading to the following linear relation
between $\dot{q}_n (u)$ and $a_\ell(u)$
\beq
\label{dqn01}
3m_0\dot{q}_n (u) = \sum_{{\rm odd\,}\ell}^{2N} \frac{a_{\ell}(u)}{\ell+n+1}.
\eeq
The right-handed side of (\ref{dqn01}) has exactly $N$ terms, implying, therefore, that one
can have at most $N$ linear independent equations of the type (\ref{dqn0}).
The linear relation between $\dot{q}_n$ and $a_\ell(u)$ given by (\ref{dqn01}) involves a Hilbert-type matrix\cite{hilbert} and,
in particular,
it is always possible to find $N+1$ rational numbers $\alpha_\ell$ such that
\beq
\label{alpha}
\frac{d}{du} \left( \sum_{\ell=1}^{N+1} \alpha_\ell q_{2\ell-1}(u) \right)= 0.
\eeq
The quantity between parenthesis is   conserved  along the solutions of (\ref{RTG}) and,
therefore, it corresponds to our second conserved quantity.
One could also truncate the summation in (\ref{dqn01}) in a given $\ell$, obtaining partial
linear combination of the odd moments that are constant along the solutions of (\ref{RTG}) up
to deviations  proportional to max $|a_{\ell+2}(u)|$. The first of such partial linear combinations are
\begin{eqnarray}
 (\ell = 0) && q_1, \\
(\ell = 1) && q_1 - \frac{5}{3}q_3, \\
\label{cond1}
(\ell = 3) && q_1 -\frac{14}{3}q_3 + \frac{21}{5}q_5, \\
&& \vdots \nonumber
\end{eqnarray}
The coefficients in the above expressions and the $\alpha_\ell$ of (\ref{alpha}) can be calculated
in a straightforward way by using, for instance, Gauss elimination in (\ref{dqn01}). However,
our numerical calculations show that,
  for the typical initial data considered here, the first odd moment $q_1$ {\em dominates}
over the other ones, implying that the typical variations $(q_1(0)-q_1(\infty))/q_1(0)$
are rather small. We notice also that the typical initial data of the family (\ref{family}) considered
here has $|q_1(0)| > |q_3(0)| > |q_5(0)| > \cdots$, in agreement with the magnitude of the
odd moments for the final state, see Fig. \ref{fig2}. This situation can fail  for some
very specific initial conditions. For instance, if one has $|q_3(0)| > |q_1(0)| > |q_5(0)| >  \cdots$,
  $q_1$ will vary  considerably along the solutions of (\ref{RTG}), but the combination
given by (\ref{cond1}) will be approximately constant, and so on.
Fig. \ref{figqn}  presents
numerical evidences confirming these results.
For practical purposes, whenever  $|q_1(0)| > |q_3(0)| > |q_5(0)| > \cdots$,
one can assume that $q_1(\infty)\approx q_1(0)$ and estimate the final
recoil velocity $v$ as
\beq
\label{vfinal}
 \frac{1}{v }\left(2- \frac{1-v^2}{v}\ln\frac{1+v}{1-v}  \right)\approx - \int_{-1}^{1} \frac{x}{Q^2(0,x)}dx.
\eeq
In particular, $v$ has the opposite sign of $q_1(0)$, see Fig. \ref{figqn}.
We emphasize, nevertheless, that (\ref{vfinal}) will be accurate solely in the
cases where $q_1(u)$ is actually the dominant moment.

\section{Discussion}

The physical properties   of the initial conditions corresponding to the family (\ref{family}) can be
investigated   by considering
their radiation content, which is determined  by  the $(1/r)$-decaying  part
of the Riemann tensor and is proportional to the quantity\cite{book, rio2}
\beq
D(u,x) = - (1-x^2) Q^2 \partial_u \left( \frac{Q_{xx}}{Q} \right).
\eeq
With the help of (\ref{RT1}) and (\ref{lambdax}), one can show that
for polynomials $Q(u,x)$ in $x$,
the function $D(u,x)$ will be also   polynomial in $x$. Moreover, $D(u,x)$
is an even (reflection-symmetric) function for even $Q(u,x)$.
The dashed lines in
Fig. 3 are polar plots without scale of $|D(0,x)|$  corresponding
to the radiation content of the associated initial data. The asymmetry in the
gravitational radiation emission  responsible for the final recoil is clear.
We notice that initial data with larger $\max |D(0,x)|$ will typically require
a larger value of the truncation order
 $N$ to attain a given accuracy in the Galerkin approximation. For instance,
 case (c) of Fig. 3 requires a truncation order larger than cases (b) and (a)
 to keep the same accuracy.

Some cases of the family (\ref{family}) are specially
interesting since they are good approximations for the Brill-Lindquist
 initial data\cite{Brill}
\beq
\label{2bh}
Q(0,x) = Q_0\left( \frac{1}{\sqrt{1 -wx}} + \frac{\mu}{\sqrt{1 +wx}} \right)^{-2},
\eeq
which can be interpreted as the final stage (after the horizon merging)
of a frontal collision of two black holes\cite{headone,headon}, with the parameters
$\mu\ge 0$ and $0\le w<1$ related, respectively, to the mass ratio and to the   infalling relative velocity
 of the two black-holes. The constant $Q_0$ must be chosen in order to assure $q_0=2$. We have
 \beq
 \label{headQ}
 Q_0^2(\mu,w) = \frac{1+\mu^4}{1-w^2} +  \frac{4\mu(1 + \mu^2)}{\sqrt{1-w^2}} + 3\frac{\mu^2}{w}
 \ln \left(\frac{1+w}{1-w} \right).
 \eeq
For $\mu=1$ (the equal masses case), the function (\ref{2bh}) is reflection-symmetric and, in this case,
the   final state of the evolution is completely determined by the constraint $q_0=2$.
For $\mu \ne 1$, one can estimate the recoil velocity for this head-on collision approximation
by using (\ref{vfinal}). For the the initial data (\ref{2bh}) we have
\begin{widetext}
\beq
\label{vxw}
\frac{1}{v }\left(2- \frac{1-v^2}{v}\ln\frac{1+v}{1-v}  \right) \approx -q_1(0)=
\frac{\left(  \mu^{-2}-\mu^2 \right)\left(
\frac{1}{w^2}\ln\frac{1+w}{1-w} - \frac{2}{w-w^3}
\right)
+ 8\left( \mu^{-1} -\mu\right) \left(
\frac{\arcsin w}{w^2} - \frac{1}{w\sqrt{1-w^2}}
\right)
}{\frac{\mu^2 + \mu^{-2}}{1-w^2} +  \frac{4(\mu + \mu^{-1})}{\sqrt{1-w^2}} + \frac{3}{w}
 \ln \left(\frac{1+w}{1-w} \right)}.
\eeq
\end{widetext}
For small values of $w$, the condition (\ref{vxw}) reduces to
\beq
v = \frac{\mu-1}{\mu+1}w.
\eeq
Fig. 6 shows the dependence of $v$ with $w$  for some values of $\mu$
 \begin{figure}[ht]
\resizebox{0.95\linewidth}{!}{\rotatebox{0}{\includegraphics*{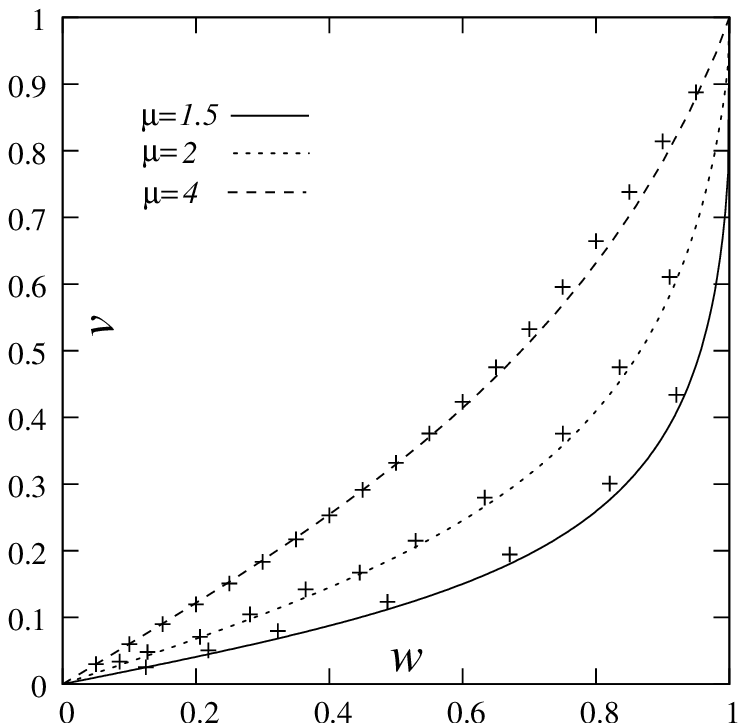}}}
\caption{ Dependence of the recoil velocity $v$ with the infalling velocity
$w$ of the two black holes with different masses, as predicted by (\ref{vxw}), and
some   results from numerical calculations.
Notice that $v\rightarrow -v$ if $\mu \rightarrow 1/\mu$.
}
\end{figure}
as predicted by (\ref{vxw}) and some numerical results. A very good agreement is
found again.  It is interesting to notice that
\beq
\lim_{w\rightarrow 1} q_1(0) = 2\frac{\mu^4-1}{\mu^4+1}
\eeq
for the initial data (\ref{2bh}), implying from (\ref{vxw}) that there exists
a maximum recoil velocity for this configuration
\beq
\label{maxvel}
\lim_{w\rightarrow 1} |v| = v_{\rm max} < 1.
\eeq
In fact, Eq. (\ref{vxw}) implies that $v<w$ for any $\mu$ (see Fig. 6), a behavior
already noticed in the numerical analysis of \cite{headon}.
One can also calculate  the
fraction (\ref{delta11}) of the initial mass radiated away  for this case
\beq
\label{delta2}
\Delta = 1  - \frac{2}{\sqrt{1-v^2}}\frac{Q_0^3(u,w)}{h(u,w)
 },
\eeq
where $v$ is given by (\ref{vxw}) and
\begin{widetext}
\beq
h(u,w) =
 \frac{2(1+\mu^6)}{(1-w^2)^2} + \frac{8(\mu+\mu^5)}{(1-w^2)^\frac{3}{2}} + \frac{15(u^2+u^4)}{1-w^2} +
 \frac{4(\mu+\mu^5)+40\mu^3}{\sqrt{1-w^2} } + \frac{15(\mu^2+\mu^4)}{2w}\ln\left(\frac{1+w}{1-w} \right).
\eeq
\end{widetext}
The aspect of the curves (\ref{delta2})   are similar to that one depicted in
Fig. 1.
In
particular, for
small $w$, one has
\beq
\Delta = \frac{3}{5}\frac{\mu(5\mu^2-8\mu+5)}{ (\mu+1)^4}w^4 + O(w^6),
\eeq
compare with (\ref{decay}). Due to (\ref{maxvel}), one has $\Delta\rightarrow 0$ irrespective of
$\mu$ for $w\rightarrow 1$.

We finish by commenting that the
  non-axisymmetric case $Q=Q(u,\theta,\phi)$ can also be investigated by means of a Galerkin method.
For such a case, the Galerkin decomposition (\ref{gal}) is based on the spherical harmonics
\beq
Q(u,\theta,\phi) = \sum_{\ell=0}^N \sum_{m=-\ell}^{\ell} b_{\ell m}(u)Y_\ell^m(\theta,\phi),
\eeq
and a system of equations equivalent to (\ref{RTG}) can be obtained. In this case,
the stationary regime corresponds
also to the case for that $b_{\ell m} = 0$ for $\ell >1$. The constant-$K$
final state will have the form
\beq
Q(\infty,\theta,\phi)  = b_{00} + b_{10}\cos\theta + a\sin\theta\cos\phi + c\sin\theta\sin\phi,
\eeq
where $b_{00}^2 - (b_{10}^2 + a^2 + c^2)=1$
for the normalized case.
The non-vanishing coefficients now can determine  the modulus and the direction of the
 Bondi's four-momentum and, consequently, the recoil velocity of the
 remnant. These topics are now under investigation.

\acknowledgements
The authors are grateful to I.D. Soares, H. Oliveira, and R. Mosna
for enlightening discussions.
This work was supported by FAPESP and CNPq.

\appendix

\section{}

One can check easily by a direct substitution that $Q(x) = a + bx$, with $a^2-b^2=K$, is a regular
solution of (\ref{lambda}).
Nevertheless, a stronger result holds in this case: {\em all} regular solutions of (\ref{lambda})
 with constant $K$
are necessarily
 of this form.
We are interested in the geometrically regular solutions ($0<Q(x)<\infty$ and
$|Q_x(x)|<\infty$
 for $-1\le x\le 1$).
The relevant phase space is three-dimensional and spanned by $(x,Q,Q_x)$. Notice that any
solution  such that $Q_{xx}=0$ must be constrained on the surface ${\cal L}$ of the phase space corresponding
to the points such that
\beq
L(x,Q,Q_x) = Q^2 - 2xQQ_x - (1-x^2)Q_x^2   - K = 0.
\eeq
One can show that, along any solution  of (\ref{lambda}), one has
\beq
\label{eqa1}
(1-x^2)Q\frac{dL}{dx} =  2(xQ + (1-x^2)Q_x)L,
\eeq
confirming that  ${\cal L}$
is indeed an invariant surface of (\ref{lambda}). The linear equation (\ref{eqa1}) has the
solution
\beq
L( x,Q(x),Q_x(x)) = A \frac{Q^2(x)}{1-x^2},
\eeq
where $A$ is a constant,  implying  that any solution of (\ref{lambda}) such that $L\ne 0$
cannot be regular in $x=\pm 1$

\end{document}